\documentclass{emulateapj}
\usepackage{graphicx}

\def\be{\begin{equation}}
\def\ee{\end{equation}}
\def\ba{\begin{eqnarray}}
\def\ea{\end{eqnarray}}
\def\go{\mathrel{\raise.3ex\hbox{$>$}\mkern-14mu
             \lower0.6ex\hbox{$\sim$}}}
\def\lo{\mathrel{\raise.3ex\hbox{$<$}\mkern-14mu
             \lower0.6ex\hbox{$\sim$}}}
\def\etal{et al.\ \rm}

\begin{document}

\title{Microwave emission from spinning dust in circumstellar disks.}
\author{Roman R. Rafikov\altaffilmark{1,2}}
\altaffiltext{1}{CITA, McLennan Physics Labs, 60 St. George St., 
University of Toronto, Toronto ON M5S 3H8 Canada; rrr@cita.utoronto.ca}
\altaffiltext{2}{Canada Research Chair}


\begin{abstract}
In the high density environments of circumstellar disks dust grains 
are expected to grow to large sizes by coagulation. Somewhat
unexpectedly, recent near-IR observations of PAH features from 
disks around Herbig Ae/Be stars demonstrate that substantial 
amount of dust mass in these disks (up to several tens of per cent of the total 
carbon content) can be locked up in particles with sizes ranging
from several to tens of nanometers. We investigate the possibility of
detecting the electric dipole emission produced by these nanoparticles 
as they spin at thermal rates (tens of GHz) in cold 
gas. We show that such emission peaks in the 
microwave range and dominates over the thermal disk emission 
at $\nu\lesssim 50$ GHz typically by a factor of several if 
$\gtrsim 5\%$ of the total carbon abundance is locked up in 
nanoparticles. We test the sensitivity of this prediction
to various stellar and disk parameters and show that if the 
potential contamination of the spinning dust component by the 
free-free and/or synchrotron emission can be removed,
then the best chances of detecting this emission 
would be in disks with small opacity, having
SEDs with steep sub-mm slopes (which minimizes
thermal disk emission at GHz frequencies). Detection 
of the spinning dust emission would provide
important evidence for the existence, properties, and origin
of the population of small dust particles in protoplanetary disks,
with possible ramifications for planet formation. 
\end{abstract}



\section{Introduction}
\label{sect:intro}


Recent IR and sub-mm observations of disks 
around young stars strongly support the idea that dust grains in 
these disks are significantly evolved compared to their ISM 
counterparts. There is mounting evidence that
in many young stellar systems dust grains have sizes reaching cm
(Testi \etal 2003; Acke \etal 2004; Natta \etal 2004), 
in contrast to sub-micron sizes of the ISM dust particles, 
which is a natural consequence 
of grain growth by coagulation. This process is 
often accompanied by the change of the material properties of dust 
grains indicated by their increased chrystallinity 
(van Boekel \etal 2005) and by settling of dust towards the 
disk midplane inferred from the change of the shape of the spectral 
energy distribution (SED) of disks (Acke \etal 2004). 
Observational evidence for dust
growth exists not only in protoplanetary disks around 
T Tauri stars (stellar mass $M_\star\lesssim 2~M_\odot$), 
but also in their analogs around higher mass 
($2~M_\odot\lesssim M_\star\lesssim 10~M_\odot$)
Herbig Ae/Be stars (van Kerckhoven, Tielens, \& Waelkens 2002; 
Sloan \etal 2005; Sch\"utz, Meeus, \& Sterzik 2005; Habart \etal 2005), 
as well as in disks around substellar objects (Apai \etal 2005).
 
At the same time, 
recent observations of the polycyclic aromatic hydrocarbon (PAH) 
spectral features in disks of Herbig Ae/Be stars (Meeus \etal 2001;
Acke \& van den Ancker 2004) provide 
evidence for the presence of a significant 
population of very small dust grains (sizes $a\sim 3-100$ \AA)
in these disks. 
Previously, the existence of a significant amount of 
carbonaceous nanoparticles in the ISM has been proposed 
to explain the {\it IRAS} observations of ``unidentified 
infrared'' emission features and strong mid-infrared emission 
component resulting from the starlight reprocessing by the 
ultrasmall grains  (Boulanger \& P\'erault 1988). 
The proposed fraction of the ISM carbon mass locked up in very 
small grains considerably exceeds that implied by the 
extension of the conventional MRN 
dust size distribution (Mathis, Rumpl, \& Nordsieck 1977) 
to very small sizes, below $50$ \AA. This suggests the existence 
of a separate population of very small carbonaceous grains 
which is distinct from the MRN distribution and contains 
$\sim 10\%$ of C in the ISM (Leger \& Puget 1984; Draine 
\& Anderson 1985). Thus, it is not surprising 
that nanoscale dust particles should exist at some level 
in disks\footnote{Possible origin of observed PAH emission in the 
roughly spherical envelopes around stars rather than in disks is
unlikely given the correlation between the strength of the PAH 
emission and the disk geometry inferred from the SED shape
(Habart \etal 2004).} around Herbig Ae/Be stars since the latter were 
originally part of the ISM. However, the detailed modeling of 
the IR emission features in these systems suggests that  
PAHs may contain as much as several tens of per cent of 
total C mass in disks (Habart, Natta \& Kr\"{u}gel 2004). 
Thus, it appear that small dust population in these 
disks does not only withstand coagulation, 
but on the contrary may have gained 
some additional mass, presumably due to the fragmentation of larger
particles. This suggests very interesting size-dependent 
evolution of the dust population in Herbig Ae/Be
disks. In the case of disks around T Tauri stars and brown dwarfs it 
is difficult to make definite statements about the presence
of small dust grains because of the lack 
of strong stellar UV fluxes needed for exciting 
PAH molecules (although see G\"urtler \etal 1999). 

In this paper we describe a new observational channel for probing 
the presence and properties of very small dust grains in 
disks around young stellar objects. In the 
outer parts of these disks, $\sim 10^2$ AU from the 
central star, nanometer-size dust particles spin at thermal rates 
of several tens of GHz. Because of their intrinsic dipole moments,
these ``macromolecules'' emit electric dipole radiation in
the microwave band where the thermal Rayleigh-Jeans emission of the 
circumstellar disk is rather weak. This makes it possible to disentangle the
spinning dust contribution to the disk emission and study the
properties of this dust component. We describe the 
physics of the spinning dust emission and analyze its observational
signatures in \S \ref{sect:dust_emission}. Applications of
our results are discussed in \S \ref{sect:disc}.


\section{Emission from spinning dust grains.}
\label{sect:dust_emission}


Ferrara \& Dettmar (1994) proposed that charged ISM dust 
grains spinning at thermal rates can produce 
observable microwave emission via electric dipole radiation. 
This mechanism has been further investigated in detail by 
Draine and Lazarian (1998b; hereafter DL98) who considered, 
among other things, intrinsic dipole moments of grains,
various processes governing the nonthermal spin rates of 
grains, and a size distribution of emitting 
nanoscale particles consistent with previous detections 
of PAH emission from the ISM. In this paper we will 
largely follow the approach adopted in these pioneering
investigations.


\subsection{Circumstellar disk model.}
\label{subsect:disk_model}


We describe the structure of the circumstellar disk using 
conventional two-temperature flared disk model of 
Chiang \& Goldreich (1997) which states 
that the vertical extent of the disk can be split into two 
well-defined regions: upper {\it exposed layer} 
in which dust absorbs most of the incoming starlight 
and reradiates it in the infrared, and lower 
{\it shielded region} which is hidden 
from direct starlight and is warmed only by the
reprocessed radiation of the outer layer.  For our purposes both 
layers can be considered isothermal with different temperatures. 
Temperature of the shielded layer $T_{sh}$ is 
lower than that of the overlaying exposed layer 
by a factor of several.

The amount of mass contained in the exposed 
layer is very small: fraction of the total disk surface 
density $\Sigma$ corresponding to this layer is 
$\sim\gamma/[\kappa_P(T_\star)\Sigma]$, where $\gamma\lesssim 0.1$ is 
the disk flaring angle and $\kappa_P(T_\star)$ is the Planck 
opacity at the stellar temperature. Since 
it is likely that even in the distant parts of the disk 
$\kappa_P(T_\star)\Sigma\gg 1$, shielded region 
must contain most of the disk mass. 
Thus, if small particles are well mixed throughout 
the disk (which is a reasonable assumption for very small 
grains having settling times longer than the disk 
lifetime) most of the small dust particles must  
be shielded from the direct stellar 
illumination by the outer exposed layer. This inference 
significantly simplifies our subsequent treatment.

We assume a simple power-law profile for the disk 
surface density 
\be
\Sigma(r)=\Sigma_1r_1^{-\alpha},
\label{eq:Sigma}
\ee
where $r_1$ is the distance from the star normalized by 1 AU,
$\Sigma_1$ is the surface density at 1 AU, and $\alpha$ is 
a parameter that we can vary. Midplane temperature profile 
$T_{sh}(r)$ needed for the calculation of 
the grain spin rates is computed in Appendix 
\ref{ap:disk_temp} according to the 
prescription of the flared disk model of Chiang \& Goldreich (1997). 

At the low temperatures of interest for us opacity $\kappa_\nu$ 
is due to dust grains. 
At long wavelengths $\kappa_\nu$ typically scales as a power law 
of $\nu$ and we adopt in this study (following the notation of 
Beckwith \etal 1990)
\be
\kappa_\nu=\kappa_{12} \nu_{12}^{\beta},
\label{eq:kappa}
\ee
where $\nu_{12}=\nu/(10^{12}\mbox{Hz})$ is the radiation
frequency normalized by $10^{12}$ Hz and $\kappa_{12}$ is 
the opacity at $10^{12}$ Hz. Power law index $\beta$ can be directly 
measured from the disk SED in the submillimeter range and 
in protoplanetary disks it varies from $0.3$ to $1.5$ 
(Kitamura \etal 2002), while ISM emission is characterized 
by $\beta\approx 1.7$ (Finkbeiner, Davis, \& Schlegel 1999). 
Small values of $\beta$ in protoplanetary disks are interpreted 
as the evidence for grain growth (Natta \etal 2004; Draine 2005).

Normalization of opacity at a given frequency $\kappa_{12}$ is 
rather poorly constrained. Beckwith \etal (1990) advocate 
using $\kappa_{12}=0.1$ cm$^2$ g$^{-1}$ while Kramer \etal (1998)
find $\kappa(1.2~\mbox{mm})\approx 0.004$ cm$^2$ g$^{-1}$
which translates into $\kappa_{12}\approx 0.016$ cm$^2$ g$^{-1}$
for $\beta=1$. Draine (2005) modeled absorbing properties 
of evolved dust populations composed of different materials 
with MRN size distributions extending to maximum size $a_{max}\sim 1-10$ cm 
and found $\beta\approx 1-1.5$ and $\kappa_{12}\sim 0.01-0.1$ cm$^2$ 
g$^{-1}$ (for the dust-to-gas ratio $10^{-2}$). Thus, there is 
at least an order of magnitude spread in the possible values 
of $\kappa_{12}$.


\subsection{Spinning dust emissivity.}
\label{subsect:emissivity}


Midplane regions of circumstellar disks have 
very low degree of ionization because of intrinsic low
temperature and strong shielding of the ionizing stellar radiation 
and cosmic ray fluxes by the overlying disk layers. As a result, 
dust grains in the shielded layer are predominantly neutral 
and  their dipole moments 
are intrinsic and not due to an asymmetric charge distribution. 
Following DL98 we assume very small dust grains to be 
composed of randomly oriented chemical substructures so that 
the total intrinsic dipole moment of the grain scales 
with the number of atoms in the grain $N$ as
\be
d\approx N^{1/2}d_0.
\label{eq:dipole_moment}
\ee
Based on the available laboratory data on the dipole moments of
the PAH-like particles DL98 have chosen $d_0=0.4$ debye in their 
study, and we adopt this estimate as well.

The power emitted by a grain of radius $a$ spinning at
frequency $\omega$ is given by 
\be
P(a,\omega)=\frac{4}{9}\frac{d^2(a)\omega^4}{c^3}=\frac{16\pi}{27}
\frac{d_0^2\omega^4}{c^3}\frac{\rho a^3}{\mu_d},
\label{eq:power}
\ee
where 
$\rho$ is the bulk density of grain material and
$\mu_d$ is the mean mass of grain atom ($\mu_d\approx 9.25$ 
amu for C:H=3:1 carbonaceous material typical for PAHs). 
In deriving equation (\ref{eq:power}) we assumed 
that grains are spherical [so that $d^2(a)=d_0^2(4\pi\rho a^3)/(3\mu_d)$] 
and that their dipole moments are randomly 
oriented with respect to their rotational axes.

To calculate the spectrum of the dipole emission of spinning
grains we also need to know the size distribution of nanoscale 
dust particles. Contrary to the expectation of small dust
depletion\footnote{DL98 assumed that
in dense interstellar clouds PAHs are depleted by a factor of $5$.} 
in circumstellar disks,
recent observations of PAH features in Herbig Ae/Be disks 
suggest that very small grains may actually 
be more abundant in these systems than in the ISM 
(Habart \etal 2004). By analogy with DL98, 
we use the following functional form for the 
grain size distribution 
\be
\frac{1}{n_H}\frac{dn}{da}=A_{\rm MRN}a^{-3.5}+\frac{B}{a}
\exp\left\{-\frac{1}{2}\left[\frac{\ln(a/a_0)}{\sigma}\right]^2
\right\},
\label{eq:distr}
\ee
where $A_{\rm MRN}=6.9\times 10^{-26}$ cm$^{2.5}$ (Draine \& Lee
1984) and $B$ are constants, $a_0$ is the typical size of 
very small carbonaceous grains, and $\sigma$ is the width of
their log-normal distribution. The first term is a conventional 
MRN contribution while the second accounts for 
very small grains. Coefficient $B$ is chosen in such a way that the
fraction of C locked up in very small grains (with
respect to the total abundance C/H=$4\times10^{-4}$) is 
$f_C$. Our baseline model uses $f_C=0.05$ (close to the ISM value), 
$a_0=3$ \AA ~and $\sigma=0.5$, this distribution  
extends down to a minimum size $a_{min}=3.5$ \AA.


\subsection{Grain spin rates.}
\label{subsect:rotation}


Rotation rates of small dust grains are in general governed by
a combination of physical processes: collisions of grains
with neutral molecules, Coulomb interactions with charged
species, infrared and electric dipole emission, etc.  
(DL98). Since here we are mainly 
interested in obtaining reasonable estimate of the importance 
of the spinning dust emission in disks, we  
assume that grain rotation is determined by thermal equilibrium. 

This assumption is reasonable in
the midplane region of the circumstellar disk because of the high
gas density and very low ionization fraction there. 
Consequently, interactions of
predominantly neutral dust grains with extremely rare charged 
particles (plasma drag/excitation and rotational excitation 
by ion collisions, which are often the most important 
determinants of grain spin rates in the ISM, see the case 
of molecular cloud environment in DL98) are completely 
negligible in this part of disk. 
As a result, grain-neutral collisions bring particle spin rates
in thermal equilibrium with the surrounding gas\footnote{Infrared 
emission by grains is unlikely to strongly affect their spin rates  
because of low midplane temperatures.}. 
This implies that grain spin rates have Boltzmann distribution 
\be
f_\omega(\omega)=4\pi\left(\frac{3}{2\pi}\right)^{3/2}
\frac{\omega^2}{\langle\omega^2\rangle^{3/2}}\exp\left(-\frac{3}{2}
\frac{\omega^2}{\langle\omega^2\rangle}\right),
\label{eq:omega_distr}
\ee
with the rms rotation rate $\langle\omega^2\rangle^{1/2}$ given by 
\be
\langle\omega^2\rangle^{1/2}=\left(\frac{3k_B T}{I}\right)^{1/2}
\approx 3.5\times 10^{10}~T_2^{1/2}
a_{-7}^{5/2}~\mbox{s}^{-1},
\label{eq:thermal_rate}
\ee 
where $T_2$ is the gas temperature normalized
to $10^2$ K, $a_{-7}$ is the grain radius in units of 
$10^{-7}$ cm, and we take $\rho=2$ g 
cm$^{-3}$. For simplicity, we assume grain shapes 
to be close to spherical implying that their moment of inertia 
is $I=(8\pi/15)\rho a^5$, although very 
small particles, like PAHs, are likely to be sheetlike (see DL98 for 
more refined treatment). 

Within the upper exposed  layer 
grains may spin at significantly non-thermal rates (corresponding
to temperatures higher than the midplane temperature) but 
the amount of mass contained in this layer 
is very small meaning that its contribution  
to the total spinning dust emission can be neglected in 
the first approximation.


\subsection{Spectrum of the spinning dust emission.}
\label{subsect:spectrum}


\begin{figure}
\plotone{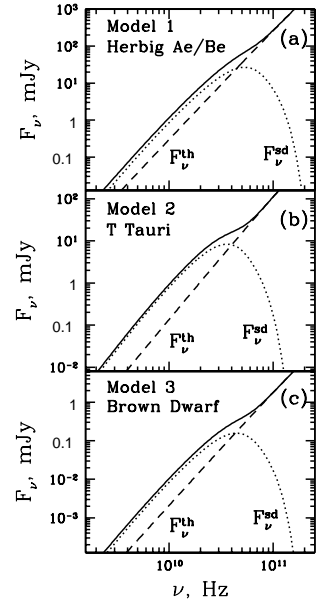}
\caption{Spectrum of the microwave disk emission for systems described
by Models 1 ({\it a}), 
2 ({\it b}), and  3 ({\it c}) placed $100$ pc away from us. Rayleigh-Jeans
tail of the thermal disk emission ({\it dashed line}), contribution of the
spinning dust emission ({\it dotted line}), and their sum ({\it solid line})
are displayed. Thermal emission is calculated for the dust emissivity index $\beta=1$
and behaves as $F_\nu^{th}\propto \nu^{3}$ (see text for other parameters). 
At radio wavelengths the spinning dust
emission exceeds the thermal disk emission by $F_\nu^{sd}/F_\nu^{th}
\approx 3$ ({\it a}), 
$\approx 6$ ({\it b}), $\approx 4$ ({\it c}).
\label{fig:models}}
\end{figure}

Combining equations (\ref{eq:power}), (\ref{eq:distr}), and 
(\ref{eq:omega_distr}) we find the total spectrum of the spinning 
dust emission produced by the disk: 
\ba
F_\nu^{sd} &=& 2\pi\int\limits_{a_{min}}^{a_{max}} da 
\frac{1}{n_H}\frac{dn}{da}\nonumber\\
& \times & \int\limits_{R_{in}}^{R_{out}}
\frac{\Sigma(r)rdr}{m_H}P(a,\omega)\times 2\pi f_\omega(\omega)\nonumber\\
&=& 8\pi^2\left(\frac{8}{3\pi}\right)^{1/2}
\frac{\omega^6}{m_H c^3}\int\limits_{a_{min}}^{a_{max}} 
\frac{1}{n_H}\frac{dn}{da}d^2(a)da \nonumber\\
& \times &  \int\limits_{R_{in}}^{R_{out}}
\frac{\Sigma(r)rdr}{\langle\omega^2\rangle^{3/2}}
\exp\left(-\frac{3}{2}\frac{\omega^2}{\langle\omega^2\rangle}\right),
\label{eq:spectrum}
\ea
where  $R_{in}$ and $R_{out}$ are the inner and outer radii of the disk, 
$\langle\omega^2\rangle$ is given by equation (\ref{eq:thermal_rate}) 
as a function of $T(r)$, and the maximum grain size $a_{max}$ is
irrelevant for this calculation since $F_\nu^{sd}$ is 
dominated by small grains.

For a simple case of a power law temperature distribution  
$T(r)=T_0(r/R_\star)^{-\gamma}$ and a dust population with 
a single characteristic size $a$ one finds that 
\ba
F_\nu^{sd}\propto \nu^{3+2(\alpha-2)/\gamma},~~~\nu_{out}
\lesssim \nu\lesssim \nu_{in},
\label{eq:spectrum_pl}
\ea
where
\ba
\nu_{out,in}\sim\left(\frac{15}{4\pi\zeta}\frac{k_B T_0}{\rho a^5}\right)^{1/2}
\left(\frac{R_{out,in}}{R_\star}\right)^{-\gamma/2}
\ea
are the frequencies determining the extent of the power law segment
of $F_\nu^{sd}$. 
Outer regions of irradiated disks  
can often be well characterized by $\gamma\approx 0.5$ which 
yields slope of the power law section of 
$F_\nu^{sd}$ quite different from $2+\beta$ characteristic
for the thermal disk emission\footnote{This 
expression accounts for the fact that at long 
wavelengths $F_\nu^{th}$ is dominated by the cool shielded
region and the contribution of the warm exposed layer is small.} 
\ba
F_\nu^{th}=2\pi\int\limits_{R_{in}}^{R_{out}}
B_\nu(T(r))\psi_{sh}(T(r))rdr 
\label{eq:thermal}
\ea 
in the optically thin Rayleigh-Jeans regime 
(here $B_\nu$ is a Planck function and $\psi_{sh}$ is defined in
Appendix \ref{ap:disk_temp}): power law index of 
$F_\nu^{sd}$ is $-1$ for $\alpha=1$ and $1$ for $\alpha=3/2$). 
As a result, one can hope to detect the spinning dust emission only 
at long wavelengths where $F_\nu^{sd}$ may dominate over 
$F_\nu^{th}$.

In Figure \ref{fig:models} we plot 
spectra of the spinning dust emission 
computed according to the prescription (\ref{eq:spectrum})
for circumstellar disks with different central 
objects at different evolutionary stages -- massive Herbig Ae/Be 
star, solar-type T Tauri star,
and a brown dwarf. Adopted parameters of these systems are
listed in Table \ref{table} and are the same\footnote{Except the 
surface density normalization in the T Tauri model which in our 
case is comparable to the Minimum Mass Solar Nebula value.} 
as in  Dullemond \& Natta (2003). In all models we use
$\alpha=1$, $\beta=1$, $\kappa_{12}=0.05$ cm$^2$ g$^{-1}$ and 
adopt a baseline model of the grain size distribution (\ref{eq:distr}) 
with parameters listed in \S \ref{subsect:emissivity}.
Disks are assumed to be located $100$ pc away from us.

One can see that all three models exhibit significant excesses 
due to the spinning dust emission around $30$ GHz. This 
component exceeds the thermal power law contribution 
by a factor ranging from $3$ to $6$. Excess appears 
first at frequencies of about $50$ GHz -- typical spin 
frequency of dust grains at low temperatures, see equation 
(\ref{eq:thermal_rate}) -- and extends all the way to lower
frequencies. 


\subsection{Sensitivity to stellar and disk parameters.}
\label{subsect:sensitivity}


The appearance of the SED including the spinning dust component 
depends on a large number of stellar, dust, 
and disk parameters:
$M_\star$, $R_\star$, $\Sigma_1$, $\alpha$, $\beta$, 
$\kappa_{12}$, etc. To test sensitivity to 
different parameters we have varied them one at a time to see
how spectrum evolves with respect to a fixed fiducial 
model of the disk+star system. As a fiducial stellar model we have adopted
T Tauri model 2 but with $M_\star=1$ M$_\star$ and $a_{in}=0.3$ 
AU for the inner edge of the disk. Fiducial model also uses
$\alpha=1$, $\beta=1$, $\kappa_{12}=0.05$ cm$^2$ g$^{-1}$. Parameters of 
the fiducial dust model are specified in \S \ref{subsect:emissivity}.

Figure \ref{fig:spec_var} demonstrates how the total microwave spectrum
of the disk $F_\nu^{sd}+F_\nu^{th}$ changes due to the parameter variation. 
Figure \ref{fig:rat_var} shows the same but
for the ratio of fluxes $F_\nu^{sd}/F_\nu^{th}$. 
These plots demonstrate that the microwave spectrum is insensitive to either 
stellar mass $M_\star$ or $R_{in}$. The former appears only 
in the calculation of the radial temperature profile (see 
Appendix A) which depends on $M_\star$ only weakly. The inner
radius of the disk does not play any role since at cm wavelengths
both the spinning dust emission and the thermal emission of
the disk are dominated by the outer region, $r\sim R_{out}$.
Variation of $R_{out}$ changes mass of the emitting gas which
mainly affects the overall flux level (Figure \ref{fig:spec_var}g) 
but also changes $F_\nu^{sd}/F_\nu^{th}$ somewhat 
(Figure \ref{fig:rat_var}g). The same is true for the spectral 
sensitivity to $\Sigma_1$ and $\alpha$ -- variation of both parameters 
mainly affects the emitting dust mass, so the characteristic pattern
of spectral evolution is similar to that caused by 
changing $R_{out}$ (Figures \ref{fig:spec_var}b,f and 
\ref{fig:rat_var}b,f). Spectral sensitivity to $T_\star$ and $R_\star$
is similar as both parameters determine the stellar luminosity. 
Changing the parameters of the small dust size distribution --
$a_0$, $\sigma$, and $f_C$ -- affects both shape and normalization
of the spinning dust spectrum leaving the thermal disk emission (dominated
by large grains) unaffected.

Variations
of $\kappa_{12}$ do not affect the spinning dust emission but 
change the optical depth of the disk which then affects $F_\nu^{th}$, see
Figure \ref{fig:spec_var}j. As a result, for low enough opacity 
(e.g. $\kappa_{12}=0.01$ cm$^2$ g$^{-1}$) spinning dust emission can 
dominate over the thermal disk emission by more than an order of 
magnitude at $\nu\lesssim 30$ GHz (see
Figure \ref{fig:rat_var}j). Similar effect is produced by 
the variation of the dust emissivity index $\beta$ --
it changes the thermal flux of the disk by orders
of magnitude since $F_\nu^{th}\propto \nu^{2+\beta}$, 
see Figures \ref{fig:spec_var}k and \ref{fig:rat_var}k. 

Total power produced by the disk determines whether
it can be detected in a flux limited observation. But even if 
such a detection is made, one can disentangle the spinning dust 
component from the thermal disk emission  only if 
$F_\nu^{sd}/F_\nu^{th}\gtrsim 1$. Examination of Figure 
\ref{fig:rat_var} demonstrates that this is most easily done for
small $\kappa_{12}$ and/or large $\beta$. Coagulation of dust
particles in dense circumstellar disks environments generally
tends to decrease $\beta$ (sometimes below 1), which would 
hide the spinning dust emission bump in the thermal power law 
component of the spectrum (see the curve for $\beta=0.5$ in Figure 
\ref{fig:spec_var}k). Thus, from the observational point of 
view the most favorable chances of detecting the spinning dust 
emission would be in systems having steep sub-mm slopes of the 
SED ($\beta\gtrsim 1$).


\section{Discussion.}
\label{sect:disc}


There is strong evidence for the presence of the spinning 
dust emission in the data obtained by the CMB experiments 
(Draine \& Lazarian 1998a; Finkbeiner 2004; although see Bennet
\etal 2004 for the opposite conclusion) and by the recent 
Green Bank Galactic Plane Survey (Finkbeiner, Langston, \& Minter 2004).
It appears that the microwave emission of the spinning dust 
in the ISM is an established physical phenomenon with
substantial observational support. For this reason there is little
doubt that it should also be present in the circumstellar disk 
environments and the two remaining issues would then be of
quantitative nature: what is the strength of this emission and whether 
it could be observed. Results presented in the previous section clearly 
demonstrate that the spinning dust contribution to the microwave emission
is substantial and in many cases significantly exceeds the thermal 
dust emission at $\nu\lesssim 30-50$ GHz. 

The answer to the second question strongly depends on the presence of 
other spectral contributions in the microwave band. 
These include free-free emission from the gas heated by the 
central star or by the nearby massive stars and synchrotron emission from 
jets and outflows which accompany large fraction of the young 
stellar objects. Spinning dust emission can be discriminated from these
components based on its intrinsic bell-like spectral shape since
both free-free and synchrotron are characterized by power-law spectra.
Power-law spectral tails at cm wavelengths extending sometimes for more 
than a decade in frequency have indeed been observed 
in some objects (Zapata \etal 2005; Eisner \& Carpenter 2006) 
suggesting the contamination by free-free and synchrotron emission.  
Given broad enough frequency coverage in the radio one can detect these
power law components at lower frequencies, extrapolate them to $10-50$ GHz 
and subtract from the total flux. Spinning dust emission will be present 
in the remaining flux provided that it dominates over such
power law contaminants (its detection would be facilitated by its 
intrinsic spectral shape).   

Another way of deducing possible free-free contamination  
is to look for H$\alpha$ emission from the same object since both
components originate in the same hot gas. This would give a
useful handle on the expected strength of the free-free flux at low 
frequencies and demonstrate a priori whether it can compete with
the spinning dust emission in the microwave domain. Synchrotron 
contribution can be identified based on its polarization and the 
morphology of the synchrotron emitting outflows (for which 
reasonable quality imaging would be required).
 
Provided that these difficulties can be successfully overcome, the 
potential detection (or nondetection) of the spinning dust emission 
can give us very interesting information on the small dust population 
in the circumstellar disks. There is already strong evidence for the 
existence of such population in disks around Herbig Ae/Be stars based 
on the detection of PAH 
emission features in the $6-12$ $\mu$m band (van Kerckhoven \etal 2002; 
Sh\"utz \etal 2005; Sloan \etal 2005; Habart \etal 2005) excited by 
copious UV flux of the central star. The total mass of nanoparticles 
producing near-IR emission features is very small ($10^{-4}-10^{-5}$ 
M$_\oplus$, see van Kerckhoven \etal 2002) but one has to remember 
that the favorable PAH excitation conditions exist only in the 
very thin outermost layer of the disk where the stellar UV 
flux gets fully absorbed. As a result, emission in PAH bands 
shows only the very tip of the iceberg as it traces an insignificant 
fraction of the total population of small dust. Most of the 
nanoparticles must be hidden from stellar UV near the disk midplane 
(provided that they are homogeneously distributed) and  
constitute substantial fraction of the total dust population (containing 
up to $30\%$ of the total C abundance, see Habart 
\etal 2004). As the spinning dust emission is contributed 
by all small dust particles (disk is optically thin at microwave
frequencies), its detection would give us a very useful probe of the 
total mass in nanoparticles. Combined 
with the model for the disk structure, the comparison of the near-IR
PAH observations and the spinning dust emission would provide a very 
good consistency check for the two probes of the small dust population. 

A nondetection of the spinning dust emission in Herbig Ae/Be systems 
may suggest that either (1) vertical mixing in disks around 
these stars is for some reason inhibited (which is unlikely given 
that the magnetorotational instability is expected to operate at 
least in the upper disk layers) or (2) PAH formation is
catalyzed by stellar UV in the uppermost disk layer and they are 
rapidly destroyed in its interior. Both possibilities appear 
rather exotic. Note that because of the
high stellar temperature the contamination with free-free emission 
may be especially severe in Herbig Ae/Be systems, potentially 
allowing only the upper limits to be set on the flux due to the 
spinning dust.

With only few exceptions (e.g. G\"urtler \etal 1999) mid-IR 
PAH emission has not been 
detected in disks around T Tauri stars or brown dwarves. This fact 
is consistent with the lack of strong UV flux  in this objects, which  
is necessary for efficient excitation of the PAH features (although see
Li \& Draine 2002 and Smith, Clayton, \& Valencic 2004  
for a different opinion), and should not be immediately interpreted 
as the evidence for the absence of the nanoparticle dust component
in these systems. Because of lack of the PAH emission, the microwave 
emission of spinning dust may be the only\footnote{The contribution 
of very small particles to opacity is degenerate with that of large 
grains.} way in which the small dust  could 
be probed in T Tauri and brown dwarf disks. 

Unlike the mid-IR PAH features, spinning 
dust emission reveals the whole nanoparticle 
dust population of the disk, which is important in many respects. 
Small particles affect the ionization balance within 
the disk since they dominate the dust 
surface area and are very efficient at immobilization
of free charges. This works against the operation
of the magnetorotational instability in protoplanetary disks 
(Fleming \& Stone 2003) which, in  turn, affects the sizes
and positions of the so-called ``dead zones'' of quiescent 
non-turbulent material (Gammie 1996), 
with potential consequences for the planet formation (Matsumura 
\& Pudritz 2005, 2006). Small grains also affect heating of 
gas in the upper rarefied disk layers due to the higher 
yield of electrons from PAHs (Kamp \& Dullemond 2004). 

The very existence of substantial amounts of nanoparticles is
very interesting from the evolutionary point of view since dust 
grains are expected to grow as disks age. Recent discoveries of  
PAHs in Herbig Ae/Be disks in amounts exceeding those
in the ISM (Habart \etal 2005) are at odds with this expectation 
but may be in line with the recent work of Dullemond \& Dominik 
(2005) who found that dust fragmentation, in addition to 
coagulation, has to be an important ingredient of the dust 
evolution. It may also be the case that the material properties
of small and large grains (e.g. sticking coefficient) are 
completely different leading to divergent evolutionary paths of the
two extremes of the grain size distribution. Detection of the 
spinning dust emission from circumstellar disks may 
shed enough light on the properties of the small dust 
population to resolve this puzzle.  

Calculations presented in \S \ref{sect:dust_emission} were 
intended to provide us with the rough idea for the importance 
of the spinning dust emission and as such they neglected a 
number of details. Among them are the realistic distribution 
of shapes of the nanoscale dust particles, emission of 
spinning dust in the exposed layer associated with 
the nonthermal grain rotation rates, possible ionization of 
small dust particles, and so on. In the future observational 
demands may warrant a more refined study including these details.


\section{Summary.}
\label{sect:summary}

We have studied the electric dipole emission produced by spinning dust 
grains of very small size (several to tens nm) in
circumstellar disks. At the temperatures characteristic for these 
environments spinning grains emit in the microwave range with
the peak of the emission occurring at $\sim 30-50$ 
GHz. For typical parameters of disks around Herbig Ae/Be stars, 
T Tauri stars, and brown dwarves we find spinning dust 
emission to dominate over the thermal disk emission 
by a factor of at least several at $\nu\lesssim 50$ GHz (if
nanoparticles contain $\gtrsim 5\%$ of the total C abundance). 
We have studied the sensitivity of this emission component 
to various stellar, disk, and dust parameters and found 
it to be strongest for variations of the total dust opacity  
--- both normalization of $\kappa_\nu$ and the slope of its 
frequency dependence. Provided 
that the contamination of the spinning dust emission by the free-free
and/or synchrotron emission can be mitigated, the best chances
of detecting this emission component should be in systems with
steep sub-mm spectral indices minimizing the 
contribution of the thermal disk emission. Detection (or 
non-detection) of the spinning dust emission will provide
important information about the existence, properties, and 
origin of the nanoscale dust particles in circumstellar 
disks.

\acknowledgements

RRR thankfully acknowledges the support of this work by 
the Canada Research Chairs program and Connaught Foundation. 

\appendix


\section{Radial temperature profile in the disk midplane.}
\label{ap:disk_temp}


Thermal balance of the shielded region is set 
by $(\delta/4) T_\star^4(R_\star/a)^2\psi_{ex}=T_{sh}^4\psi_{sh}$,
where $T_{sh}$ and $T_{ex}$ are the characteristic temperatures 
of the shielded and exposed regions respectively and $\psi_{ex}$ and $\psi_{sh}$
are the fractions of the radiative flux with blackbody temperatures 
$T_{ex}$ and $T_{sh}$ absorbed by the disk. In the optically thick
case $\psi_{ex,sh}=1$ while in the optically thin regime 
$\psi_{ex,sh}=\Sigma\kappa_P(T_{ex,sh})$. Flaring angle $\delta\equiv
d(H_1/a)/d\ln a$ is determined by the height of the disk surface $H_1
\approx \lambda c_s(T_{sh})/\Omega$, where $c_s$ is the gas sound 
speed and $\Omega$ is the angular frequency of the disk. Factor 
$\lambda\sim 1$ is roughly constant throughout the disk
and we set $\lambda=3$ in this study. Power law scaling of $\kappa_\nu$
implied by equation (\ref{eq:kappa}) results in the following expression 
for the low temperature Planck opacity: 
\ba
\kappa_P(T)=\kappa_\star\left(\frac{T}{T_\star}\right)^\beta,~~~\kappa_\star=
\frac{15}{\pi^4}\frac{\kappa_{12}}{(10^{12}\mbox{Hz})^{\beta}}
\left(\frac{k_B}{h}\right)^\beta\int\limits_0^\infty\frac{x^{3+\beta}dx}{e^x-1},
\label{eq:kappa_P_pl}
\ea
where $\beta$ is the same as the power law index in equation 
(\ref{eq:kappa}). Opacity dependence given by (\ref{eq:kappa_P_pl}) applies
to the emission of both the disk interior and the exposed layer,
but not to the stellar radiation, so that $\kappa_P(T_\star)$ is significantly
different from $\kappa_\star$. In our case $\kappa_P(T_\star)$ is calculated 
for a population of small $0.1~\mu$m dust grains (Draine \& Lee 1984; Dullemond 
\& Natta 2003). Using (\ref{eq:kappa_P_pl}) one finds that
\ba
T_{ex}(r)=T_\star\left[\frac{\kappa_P(T_\star)}{4\kappa_\star}\right]^{1/(4+\beta)}
\left(\frac{R_\star}{r}\right)^{2/(4+\beta)}
\label{eq:t_ex}
\ea

This information allows one to derive the temperature profile 
in the shielded layer in different parts of the disk. Inner 
disk is optically thick to radiation at both $T_{ex}$ and $T_{sh}$,
which results in 
\ba
T_{sh}(r)=T_\star\left(\frac{\lambda}{14}\frac{h_\star}{R_\star}\right)^{2/7}
\left(\frac{r}{R_\star}\right)^{-3/7}.
\label{eq:T1}
\ea
In the intermediate region disk is optically thin to the radiation 
of its interior [$\psi_{sh}=\Sigma\kappa_P(T_{sh})<1$] while 
still optically thick to the reprocessed emission of the exposed 
layer ($\psi_{ex}=1$). As a result,
\ba
T_{sh}(r)=T_\star\left[\frac{2+\alpha+\beta}{4(7+2\beta)}\frac{\lambda}
{\kappa_\star\Sigma_\star}\frac{h_\star}{R_\star}\right]^{2/(7+2\beta)}
\left(\frac{r}{R_\star}\right)^{(2\alpha-3)/(7+2\beta)},
\label{eq:T2}
\ea
where $\Sigma_\star=\Sigma(R_\star)$ is a  value of $\Sigma$ obtained by  
extrapolation of (\ref{eq:Sigma}) to $r=R_\star$. Finally, in 
the outer disk, which is optically thin to the radiation at both
$T_{ex}$ and $T_{sh}$, one finds
\ba
T_{sh}(r)=T_\star\left[\frac{\beta^2+4\beta+8}{4(4+\beta)(7+2\beta)}\lambda
\frac{h_\star}{R_\star}\right]^{2/(7+2\beta)}
\left[\frac{\kappa_P(T_\star)}{4\kappa_\star}\right]^{2\beta/[(4+\beta)(7+2\beta)]}
\left(\frac{r}{R_\star}\right)^{-(7\beta+12)/[(4+\beta)(7+2\beta)]},
\label{eq:T3}
\ea
Transition between the inner and intermediate regions occurs at
\ba
a_1\approx R_\star\left(\kappa_\star\Sigma_\star\right)^{7/(7\alpha+3\beta)}
\left(\frac{\lambda}{14}\frac{h_\star}{R_\star}\right)^{2\beta/(7\alpha+3\beta)}
\label{eq:a1}
\ea
and between the intermediate and outer regions at
\ba
a_2\approx R_\star\left(\kappa_\star\Sigma_\star\right)^{(4+\beta)/[2\beta+\alpha(4+\beta)]}
\left[\frac{\kappa_P(T_\star)}{4\kappa_\star}\right]^{\beta/[2\beta+\alpha(4+\beta)]}.
\label{eq:a2}
\ea
In a particular case of $\alpha=3/2$, $\beta=1$ 
our results agree with Chiang \& Goldreich (1997). The profile of
$T_{sh}(r)$ used in  \S \ref{sect:dust_emission} is found by 
extrapolation between expressions (\ref{eq:T1})-(\ref{eq:T3}).


\begin{center}
\begin{deluxetable}{ l l l l l l l l }
\tablewidth{0pc}
\tablecaption{List of models used for calculation of the spinning dust emission. 
\label{table}}
\tablehead{
\colhead{Model}&
\colhead{$M_\star$ $[M_\odot]$}&
\colhead{$R_\star$ $[R_\odot]$}&
\colhead{$T_\star$ [K]}&
\colhead{$R_{in}$ [AU]}&
\colhead{$R_{out}$ [AU]}&
\colhead{$\Sigma_1$ [g cm$^{-2}$]}&
\colhead{Object}
}
\startdata
1 & $2$ & $2$ & 10000 & 1 & 300 & 1000 & Herbig Ae/Be \\
2 & $0.5$ & $2$ & 4000 & 0.1 & 300 & 1000 & T Tauri \\
3 & $0.1$ & $1.3$ & 2600 & 0.033 & 30 & 100 & Brown Dwarf \\
\enddata
\end{deluxetable}
\end{center}

\begin{figure}
\plotone{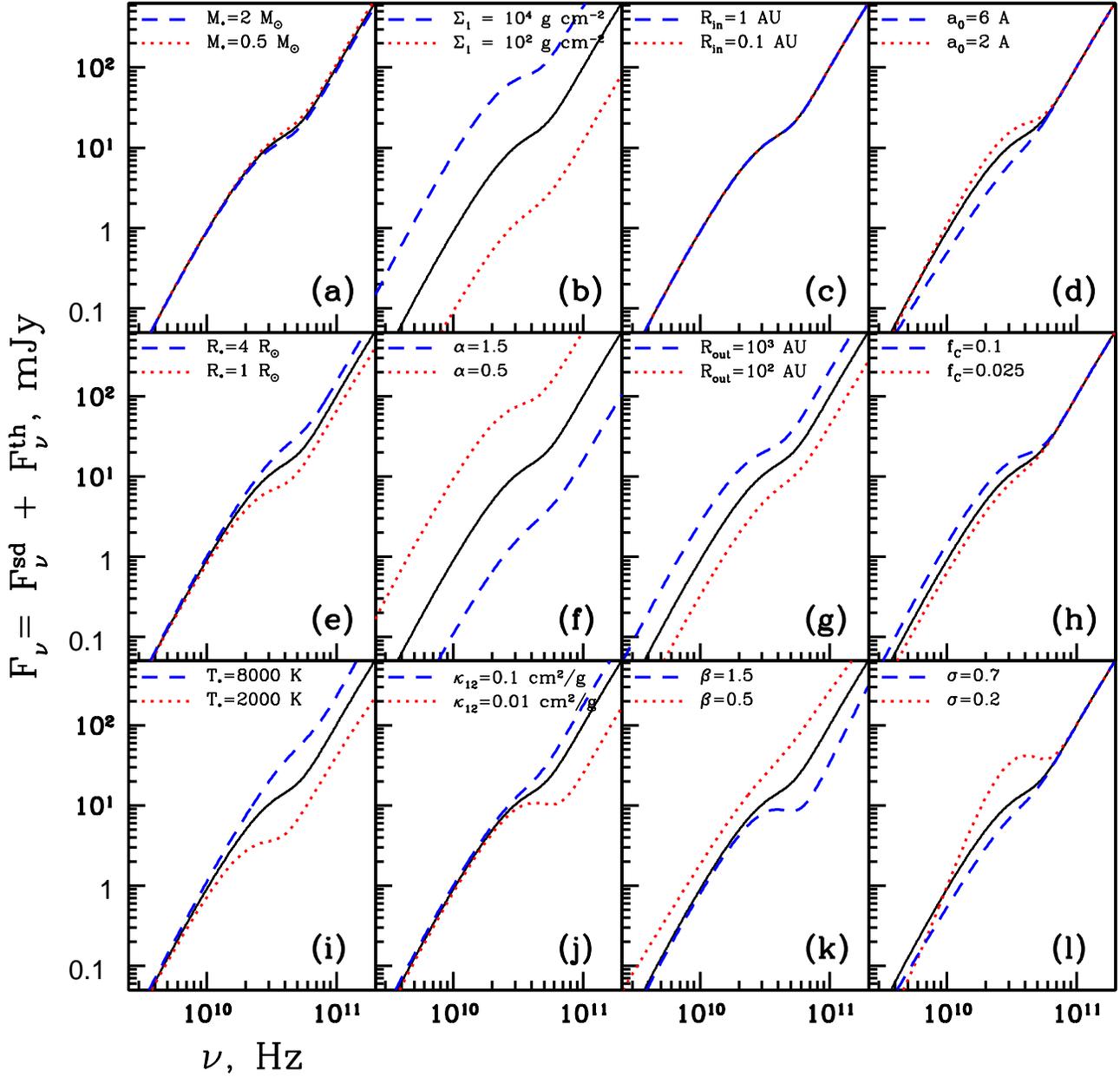}
\caption{Evolution of the total disk spectrum (both thermal 
and spinning dust contributions included) caused by the 
variation of basic model parameters. Each window displays 
the results obtained by variation of a single parameter around
the fiducial model described in the text. The parameter being 
varied and its values corresponding to different curves  
are indicated in each window. Solid curve
always represents the spectrum of the fiducial model. 
\label{fig:spec_var}}
\end{figure}

\begin{figure}
\plotone{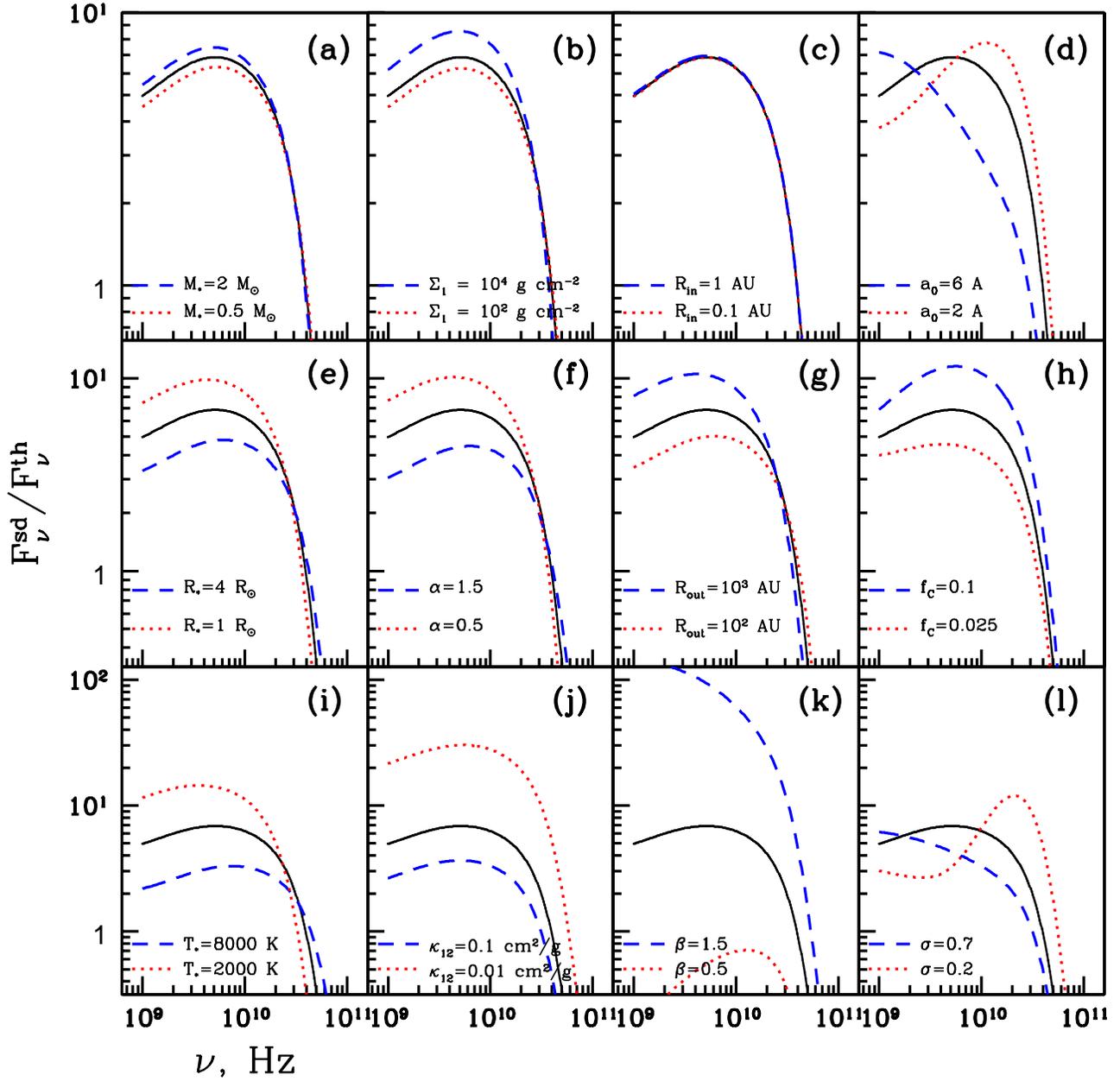}
\caption{Evolution of the ratio $F_\nu^{sd}/F_\nu^{th}$
representing the strength of the spinning dust contribution 
relative to the thermal disk emission due to the variation 
of the model parameters with respect to the fiducial model. 
Notation is the same as in Figure \ref{fig:spec_var}.
\label{fig:rat_var}}
\end{figure}

\end{document}